[1994/12/01]
\documentclass{article}

\usepackage{arxiv}

\usepackage[utf8]{inputenc} 
\usepackage[T1]{fontenc}    
\usepackage{hyperref}       
\usepackage{url}            
\usepackage{booktabs}       
\usepackage{amsfonts}       
\usepackage{nicefrac}       
\usepackage{microtype}      
\usepackage{lipsum}

\usepackage{amsmath,amsfonts,amssymb,amsbsy,algorithm,algorithmic}
\usepackage{graphicx,footnote}
\usepackage{enumerate}
\usepackage{color}
\usepackage[displaymath,mathlines]{lineno}
\usepackage{natbib}
\usepackage{etoolbox}

\title{Towards a Prototype Global CO$_2$ Emissions Monitoring System for Copernicus. \MakeUppercase{\romannumeral 1}: Methodological Aspects}
\author{Nicolas Bousserez\\
 European Center for Medium-Range Weather Forecast \\
 Research Department}
\date{June 2019}
\begin{document}

\maketitle
\begin{abstract}
This document describes in details a new hybrid ensemble-variational algorithm that generalizes existing ensemble or variational data assimilation approaches and would enable joint state/parameter estimations in ECMWF's Integrated Forecasting System (IFS). The proposed methodology is intended to serve as the basis for a prototype Copernicus CO$_2$ emission monitoring service. The main characteristics of the system are: 1) 4D hybridization of ensemble information with full-rank statistical modeling by combining an ensemble-based increment with an adjoint-based increment propagation, allowing one to increase current spatial resolution and/or include forward model processes missing from the adjoint integration; 2) combination of tangent-linear and adjoint solvers with ensemble-based approximations of transport Jacobians to construct a long-window 4D-Var with timescales relevant to greenhouse gas source inversion. The proposed methodology is non-intrusive in the sense that the main structure of the current incremental 4D-Var algorithm remains unchanged, while the additional computational cost associated with the source inversion component is minimized. 
\end{abstract}
\section{Context and rationale}
\subsection{Requirements}
\label{requirements}
The Carbon Human Emission (CHE) project aims to build a prototype global CO$_2$ source inversion system that can provide policy-relevant information on the spatio-temporal characteristics of anthropogenic CO$_2$ emissions. This prototype shall evolve toward an operational Copernicus CO$_2$ service, which will provide a Monitoring and Verification Support (MVS) capacity and will build on existing infrastructures (CAMS, C3S) to exploit ground-based measurements as well as space-based observations from current and future satellite missions (e.g., Sentinel 5p and Sentinel 7). Requirements for such as a prototype are described in details in \cite{pinty2017}. The minimum set of capabilities is defined as:
\begin{itemize}
    \item Detection of emitting hot spots such as megacities or power plants.
    \item Monitoring the hot spot emissions to assess emission reductions of the activities.
    \item Assessing emission changes against local reduction targets to monitor impacts of the Nationally Determined Contributions (NDCs).
    \item Assessing the national emissions and changes in 5-year time steps to estimate the global stock take.
\end{itemize}
In addition to the need for separate research activities to enhance our capacities within each system element (i.e., in situ and space-based observations, emission inventories and atmospheric transport modeling), an important requirement for a successful implementation of an operational CO$_2$ monitoring system is the integration of all of those elements together in a consistent and efficient manner. This includes the use of a robust methodological framework, advanced computational algorithms that ensure good scalability with respect to the increasing number of available measurements, and the generation of useful posterior emission products and their associated uncertainties to support environmental decision-making. The present document focuses on this integration activity and introduces methodological concepts that would enable:
\begin{itemize}
    \item Anthropogenic CO$_2$ source attribution through joint assimilation of atmospheric observations of CO$_2$ and co-emitted tracers (e.g., CO and NO$_2$) for CO$_2$.
    \item Fusion of heterogeneous posterior emission products (from global to regional to local) through their assimilation into a global CO$_2$ source inversion system based on ECMWF's Integrated Forecasting SYstem (IFS).
\end{itemize}

\subsection{Characteristics of the IFS system}
\label{sec:IFS_const}

In this section we describe two characteristics of the current IFS system that have implications for the implementation of a chemical source inversion system: the short 12-hour 4D-Var window used in operation and the absence of chemical reaction mechanisms in the tangent linear (TL) and adjoint (AD) models. 

The IFS uses an incremental 4D-Var assimilation system \cite{courtier1994} with a 12-hour window. Strong non-linearities associated with the dynamics of the numerical weather model prevent one to use long assimilation windows (e.g., several days to months), since this would hamper the computational efficiency of the minimization due to the presence of multiple minima. However, in the IFS the propagation of error statistics from one window to the next is ensured by using an ensemble of data assimilation (EDA) system that mimics the error propagation procedure performed in stochastic ensemble Kalman filter (EnKF) \cite{evensen1994}. In the context of a CO$_2$ source inversion prototype, given the atmospheric lifetime of CO$_2$ (5 to 200 years), the IFS would need to be adapted so as to provide Kalman smoother capabilities, i.e., the possibility to utilize observations in the current 12-hour assimilation window to constrain CO$_2$ fluxes defined in previous windows.

Another issue pertaining to atmospheric composition aspects of the IFS is that there are currently no chemical mechanisms included in the TL and AD models. As a result, the transport of chemical tracers is the only process accounted for in the quadratic minimization step (inner-loop) of incremental 4D-Var. While this simplification might be appropriate for an inert trace gas such as CO$_2$, it can result in significant errors in the DA of reactive trace gas concentrations as well as in the inversion of their precursors' emissions. The CO$_2$ source inversion prototype should accommodate reactive species and their precursors for two reasons. First, precursors emissions are the main driver of air quality forecast \cite{bocquet2015}, and therefore it is expected that improving our estimate of these sources will have a positive impact on the quality of the CAMS products. Secondly, as emphasized in \cite{pinty2017}, the potential of exploiting available and future atmospheric observations of co-emitted species (e.g., NO$_2$ and CO) to complement constraints on CO$_2$ fluxes (especially for source differentiation) should be investigated.

In order to address the two shortcomings described above (i.e., short assimilation window and transport-only AD/TL models), two adaptations of the current IFS system are proposed: one that extends the control vector to past (and future) DA windows, and one that generalizes to the time dimension the hybrid background error covariance formulation in order to exploit ensemble information that is missing from the TL/AD model integrations (e.g., chemical reactions, higher spatial resolution).

\section{Methodology}
\subsection{Hybrid ensemble-variational system}
\subsubsection{General principle}
\label{sec:hyb_B}
The four dimensional variational (4D-Var) method consists in minimizing the following cost function:
\begin{align}
\label{eq1}
J(\mathbf{x})=(\mathbf{x}-\mathbf{x}_b)^T{\mathbf{B}}^{-1}(\mathbf{x}-\mathbf{x}_b) +(\mathbf{y}-h(\mathbf{x}))^T\mathbf{R}^{-1}(\mathbf{y}-h(\mathbf{x}))  ,
\end{align}
where $\mathbf{x}$ is the model variable to be optimized, $\mathbf{x}_b$ the mean of the associated prior Gaussian distribution (usually called the background), $\mathbf{B}$ its associated error covariance matrix, $\mathbf{y}$ is the vector of observations with associated error covariance matrix $\mathbf{R}$, and $h$ is the observation operator mapping the model variable $\mathbf{x}$ onto the observations.
In the most general form, adopted here, all mathematical quantities (i.e., vectors and matrices) defined in (\ref{eq1}) are 4D objects defined in space and time, and $\mathbf{x}$ can include both model prognostic and parameter (e.g., forcing) variables. Different flavors of 4D-Var stem from the method chosen to approximate $\mathbf{B}$. For instance, in the strong-constrained 4D-Var framework, the TL and AD model integrations are used to propagate (implicitly) the initial background error statistics at $t_0$ throughout the assimilation window, that is:
\begin{align}
\label{eq2}
\forall t_k,\, \mathbf{B}(t_{k})=\mathbf{M}_{t_0 \rightarrow t_k} \mathbf{B}(t_0)\mathbf{M}_{t_k \rightarrow t_0}^T \text{ (strong-constrained 4D-Var)},
\end{align}
where $\mathbf{B}(t)$ is the background error covariance matrix of the variable $\mathbf{x}$ at time $t$, noted $\mathbf{x}(t)$ thereafter, and $\mathbf{M}_{t_0 \rightarrow t}$ and $\mathbf{M}^T_{t \rightarrow t_0}$ represent the TL and AD model integrations, respectively, between $t_0$ and $t$. Here the time discretization of the error propagation is made explicit by using the indices $k$. 
In another formulation, $\mathbf{B}$ is assumed to be constant, i.e.:
\begin{align}
\label{eq3}
\forall t_k,\, \mathbf{B}(t_k)= \mathbf{B}(t_0)\text{  (3D-FGAT)}
\end{align}
In the case of weak-constrained 4D-Var, the propagation of the background error statistics can be defined recursively as:
\begin{align}
\label{eq4}
\forall t_k,\, \mathbf{B}(t_{k+1})= \mathbf{M}_{t_{k} \rightarrow t_{k+1}} \mathbf{B}(t_{k}) \mathbf{M}_{t_{k+1} \rightarrow t_{k}}^T +\mathbf{Q}(t_{k+1}) \text{  (weak-constrained 4D-Var)},
\end{align}
where $\mathbf{Q}(t_{k+1})$ is an error covariance matrix that represents the forward model errors associated with its integration between time $t_k$ and $t_{k+1}$. Since those model error statistics $\{\mathbf{Q}(t_k),\, k=1,...,n\}$ are assumed to be uncorrelated, an efficient parallel implementation of incremental weak-constrained 4D-Var is possible, wherein $n$ quadratic minimization problems are solved independently using the model errors $\{\eta_{k}\equiv\mathbf{x}_{k+1}-M(\mathbf{x}_{k}),\, k=1,...,n\}$ as control variables. Note that in practice the model error variables $\eta_{k}$ are defined at regular time intervals that are coarser than the TL model integration time-step.

In (\ref{eq2}) and (\ref{eq4}), the propagation of the error statistics is entirely determined by the TL and AD models. Therefore, a coarser spatial resolution and/or approximations of the forward model processes in the TL and AD model integrations can significantly limit the quality of the resulting 4D background error statistics. For instance, in the case of the CAMS analysis, the TL and AD models only include transport processes, i.e., chemical reactions in the atmosphere are not accounted for in the quadratic minimizations. 

Ensemble methods have been designed to by-pass the need for TL and AD models. While the latter implicitly propagate error statistics, the former consist in explicitly propagating a sample estimate of those errors. The advantage is twofold: 1) only the forward model is necessary to propagate an ensemble of perturbed states, avoiding the need for development and maintenance of TL and AD models; 2) the ensemble of forward model integrations can be carried out in parallel, providing potential improvements in computational scalability. Another advantage of ensemble methods is their flexibility. For instance, ensembles can be generated from heterogeneous sources of information (e.g., different models) and merged together into one sample pdf. However, sample estimates of the error statistics are necessarily very low-rank compared to the dimension of the problem. While the implicit full-rank propagation of the error statistics in variational adjoint-based methods is limited by required approximations in the modeling of $\mathbf{B}$ (i.e., square-root operator, TL/AD propagation), the accuracy of ensemble-based approaches is affected by sampling noise, a problem generally known as the curse of dimensionality. The sampling noise in ensemble-based estimates is usually filtered out using localization techniques \cite{hamill2001}.

Here we describe an approach that allows one to correct for missing (or simplified) processes in the TL/AD model propagation of the background error statistics by incorporating information from an ensemble of forward model simulations. This method optimally combines the advantages of both variational and ensemble-based methods. It can be interpreted as a generalization of the so-called hybrid background error covariance approach, wherein the $\mathbf{B}$ matrix at the beginning of the 4D-Var window is defined as a weighted average of a static and an ensemble-based (flow-dependent) background error covariance matrix, as follows:
\begin{align}
\label{eq5}
\mathbf{B}(t_0)=\beta \mathbf{B}_\text{stat}(t_0)+ (1-\beta)\mathbf{C} \circ \mathbf{B}_\text{ens}(t_0),
\end{align}
where $0<\beta<1$, $t_0$ is the initial time of the window, $\mathbf{B}_\text{stat}$ is a static background error covariance modeled using a control variable transform (CVT) approach \cite{bannister2008}, $\mathbf{B}_{ens}$ represents a sample (ensemble-based) estimate of the background error statistics based on perturbed forward model simulations, and $\mathbf{C}$ is a symmetrix matrix whose Schur product ($\circ$) with $\mathbf{B}_{ens}$ is used for localization of the sampling noise. 
In practice $\mathbf{C}$ is 4D, although for the sake of simplicity in the following the localization matrix will be assumed to constant throughout the assimilation window. 
The implementation of 4D localization will be addressed in Sec. \ref{sec:loc}.

A natural generalization of (\ref{eq5}) is to define recursively the $\mathbf{B}$ matrix as:
\begin{align}
\label{eq6}
\mathbf{B}(t_{k+1})=\beta  \mathbf{M}_{t_{k} \rightarrow t_{k+1}} \mathbf{B}(t_{k}) \mathbf{M}_{t_{k+1} \rightarrow t_{k}}^T + (1-\beta) \mathbf{C} \circ \mathbf{B}_\text{ens}(t_{k+1}),
\end{align}
where $\mathbf{B}_\text{ens}(t)$ is the sample estimate of the background error covariance matrix at time $t$.
Note the difference between (\ref{eq4}) and (\ref{eq6}). While in (\ref{eq4}) the model errors between consecutive times are independent and therefore additive, in (\ref{eq6}) the ensemble-based  error statistics are in general correlated throughout the assimilation window. Finally, a more general $\mathbf{B}$ matrix that combines weak-constrained 4D-Var (i.e., model error bias) with a hybrid 4D $\mathbf{B}$ matrix can be defined as:
\begin{align}
\label{eq7}
\mathbf{B}(t_{k+1})=\beta  \mathbf{M}_{t_{k} \rightarrow t_{k+1}} \mathbf{B}(t_{k}) \mathbf{M}_{t_{k+1} \rightarrow t_k}^T + (1-\beta) \mathbf{C} \circ \mathbf{B}_\text{ens}(t_{k+1})+\mathbf{Q}(t_{k+1}),
\end{align}
where $0<\beta<1$. In that context, while $\mathbf{B}_\text{ens}$ includes model random errors within the assimilation window (e.g., through stochastic perturbations of the physical tendencies (SPPT)), $\mathbf{Q}$ is associated to the model bias within that window. Those error statistics are therefore complementary to and independent from each other. 
In the following, only the hybrid ensemble-variational formulation (\ref{eq6}) will be considered. This formulation of 4D-Var combines both ensemble-variational (EnVar) and adjoint-based (strong-constrained) characteristics. Indeed, at each time-step a full-rank approximation of $\mathbf{B}$ is propagated, but linearly combined with an low-rank ensemble-based estimate. This simultaneously addresses several problems related to the use of either approach. First, the perfect model assumption in strong-constrained 4D-Var is relaxed through the use of sampled random model errors in the background state. This limitation, although currently circumvented by the use of an EnKF-like Ensemble of Data Assimilation (EDA) method, is now addressed within one single 4D-Var minimization. Secondly, the sampling noise issue in EnVar can be at least partially mitigated via the linear combination with a full-rank propagated increment \cite{menetrier2015}.

\subsubsection{Minimization}
\label{sec:min}
In this section we describe the practical implementation of the incremental 4D-Var minimization in the framework of the hybrid ensemble-variational system described in Sec. \ref{sec:hyb_B}. In incremental 4D-Var, a series of quadratic cost functions are minimized, which correspond to the initial variational problem (\ref{eq1}) linearized around updated non-linear trajectories. Each quadratic cost function is defined as:
\begin{align}
\label{eq8}
J(\delta \mathbf{x}_0,...,\delta \mathbf{x}_K)=&\frac{1}{2}\sum_{k=1}^K(\delta\mathbf{x}_k-\mathbf{d}_k^b)^T{\mathbf{B}_k}^{-1}(\delta\mathbf{x}_k-\mathbf{d}_k^b) +\frac{1}{2}\sum_{k=1}^K(\mathbf{H}_k\delta \mathbf{x}_k-\mathbf{d}_k^o)^T\mathbf{R}_k^{-1}(\mathbf{H}_k\delta \mathbf{x}_k-\mathbf{d}_k^o) , 
\end{align}
where $k$ is a time indice associated with the discretization of the problem in ($K$) subwindows, $\mathbf{d}_k^{b}=\mathbf{x}_k^b-\mathbf{x}_k^t$ and $\mathbf{d}_k^{o}=\mathbf{y}_k-h_k(\mathbf{x}_k^t)$, where $\mathbf{x}_k^b$ and $\mathbf{x}_k^t$ represent the background and first guess trajectory, respectively. In practice, (\ref{eq8}) is often reformulated by preconditioning the problem using the square-root of $\mathbf{B}$, i.e.: 
\begin{align}
\label{eq9}
J(\mathbf{v}_0,..., \mathbf{v}_K)=&\frac{1}{2}\sum_{k=1}^{K}(\mathbf{v}_k-\mathbf{d}_k^{\mathbf{v},b})^T(\mathbf{v}_k-\mathbf{d}_k^{\mathbf{v},b})+\frac{1}{2}\sum_{k=1}^{K}(\mathbf{H}_k \mathbf{L}_k\mathbf{v}_k-\mathbf{d}_k^{o})^T\mathbf{R}_k^{-1}(\mathbf{H}_k\mathbf{L}_k\mathbf{v}_k-\mathbf{d}_k^{o}), 
\end{align}
where:
\begin{align*}
&\mathbf{B}_k =  \mathbf{L}_k\mathbf{L}_k^T   \\ 
&\delta \mathbf{x}_k =  \mathbf{L}_k \mathbf{v}_k \\ 
& \mathbf{d}_k^{\mathbf{v},b} =\mathbf{L}_k^{-1}\mathbf{d}_k^{b} 
\end{align*}
The incremental 4D-Var algorithm requires to evaluate the cost function (\ref{eq9}), its gradient as well as the product of its Hessian by a vector. The gradient and the Hessian-vector product can be expressed as:
\begin{align}
\label{eq10}
&\nabla J_{(\mathbf{v}_0,..., \mathbf{v}_K)} & =&\sum_{k=1}^{K}(\mathbf{v}_k-\mathbf{d}_k^{\mathbf{v},b})+\sum_{k=1}^{K} \mathbf{L}_k^T\mathbf{H}_k^T\mathbf{R}_k^{-1}\left ( \mathbf{H}_k\mathbf{L}_k \mathbf{v}_k - \mathbf{d}_k^{o} \right ) \\
&\nabla^2 J_{(\mathbf{v}_0,..., \mathbf{v}_K)} \mathbf{u}_k&=&\sum_{k=1}^{K}  \left (\mathbf{I}+\mathbf{L}_k^T\mathbf{H}_k^T\mathbf{R}_k^{-1} \mathbf{H}_k\mathbf{L}_k  \right ) \mathbf{u}_k , \nonumber
\end{align}
where $\mathbf{u}_k$ represents a vector defined in the same space as $\mathbf{v}_k$.

We now need to express those quantities in the case where the $\mathbf{B}$ matrix is defined by (\ref{eq6}). Note that since the ensemble-based covariances $\mathbf{B}_\text{ens}(t_k)$ are in general correlated in time (unlike in the weak-constrained case), the TL/AD propagation of the error covariances cannot be decoupled between subwindows. From (\ref{eq10}) one sees that only the square-root-vector products ($\mathbf{L}_k\mathbf{v}_k$) and their adjoint counterparts ($\mathbf{L}_k^T \mathbf{x}_k$) need to be expanded.
At any given iteration of the quadratic minimization that has produced a new increment $\mathbf{v}$, assuming the departures between the background and the first guess non-linear trajectory equivalent $\mathbf{d}_k^{b}$ and the departures between the observations and the model equivalebt $\mathbf{d}_k^{o}$ have been evaluated, and considering $p$ samples to estimate $\mathbf{B}_\text{ens}$, one can use a recursive algorithm to compute the square-root-vector product $\mathbf{L}\mathbf{v}$. Let us define:
\begin{align*}
&{\mathbf{w}_k^i}'&=&1/\sqrt{p-1}(\mathbf{w}_k^i-\overline{\mathbf{w}_k}),\,i=1,...,p,\,k=0,...,K-1  \\ 
& \mathbf{B}_\text{ens} &=&\mathbf{C}\circ\mathbf{W'}\mathbf{W'}^T,
\end{align*}
where
\[
\mathbf{w}_k^i
=
\begin{bmatrix}
w_k^i(1) \\
\vdots \\
w_k^i(n)
\end{bmatrix}
\]
is the state vector of dimension $n$ associated with the $i$th sample at time $k$, $\overline{\mathbf{w}_k}$ represents the average vector over the samples at time $k$,
\[
\mathbf{W}'
=
\begin{bmatrix}
    {\mathbf{w}_1^1}'       & {\mathbf{w}_1^2}' & \dots & {\mathbf{w}_1^p}' \\
    {\mathbf{w}_2^1}'       & {\mathbf{w}_2^2}' &  \dots &{\mathbf{w}_2^p}'\\
     \vdots & \vdots & \ddots & \vdots \\
   {\mathbf{w}_{K-1}^1}'    & {\mathbf{w}_{K-1}^2}' & \dots & {\mathbf{w}_{K-1}^p}'
\end{bmatrix}
\]
is the matrix whose columns are the sample vectors concatenated across all ($K-1$) subwindows, and $\mathbf{C}$ is a 4D localization matrix. The vector $\mathbf{w}_k^i$ represents the contribution of the $i$th ensemble vector perturbation associated with subwindow $k$ to $\mathbf{B}_\text{ens}$. Let us define the control vector:
\begin{align*}
\mathbf{v}=\left ({\mathbf{u}_0},{\mathbf{s}_i},\, i=1,...,p \right )
\end{align*}
The following algorithm computes $\mathbf{L}\mathbf{v}$:
\begin{align}
\label{eq11}
&\boldsymbol{\alpha}_i&=&\mathbf{C}^{1/2}\mathbf{s}_i,\,i=1,...,p \\ \nonumber
&\mathbf{q}_0&=&\beta_0 \mathbf{L}_0\mathbf{u}_0 + (1-\beta_0) \sum_{i=1}^p\lambda_i^{1/2} \boldsymbol{\pi_0}(\boldsymbol{\alpha}_i \circ{\mathbf{w}^i}') \\ \nonumber
&\mathbf{q}_{k+1} &= &\beta_{k+1} \mathbf{M}_{k,k+1} \mathbf{q}_{k} + (1-\beta_{k+1})\sum_{i=1}^p\lambda_i^{1/2} \boldsymbol{\pi_k}(\boldsymbol{\alpha}_i\circ{\mathbf{w}^i}'),\,k=0,...,K-1 \\ \nonumber
&\mathbf{L} \mathbf{v}&=&\left ( \mathbf{q}_{k},\,k=0,...,K \right ),
\end{align}
where $\mathbf{L}_0$ is the square-root of the background error covariance matrix at time $t_0$, which is modeled using, e.g., a CVT approach, ${\mathbf{w}^i}'$ represents the $i$th column of $\mathbf{W}'$, $0<\beta_i<1$ are scalar weights, and $\boldsymbol{\pi_k}$ represents the orthogonal projection operator onto the subspace associated with time $k$. 
A similar recursive method can be derived for the adjoint counterpart. Defining a increment: 
\begin{align*}
\mathbf{x}=\left ( \mathbf{x}_k,\,k=0,...,K   \right ) ,
\end{align*}
where the $\mathbf{x}_k$ are $n$ dimensional vectors defined at all time indices $k=0,...,K-1$, one can compute the adjoint-vector product using the following algorithm:
\begin{align}
\label{eq12}
&\mathbf{q}'_{K}&=&\mathbf{x}_K\\ \nonumber
&\mathbf{q}'_{k} &= &\beta_{k+1} \mathbf{M}^T_{k+1,k} \mathbf{q}'_{k+1}+\mathbf{x}_k,\,k=1,...,K-1 \\ \nonumber
&\boldsymbol{\alpha}'_{i} &=& \boldsymbol{\alpha}'_{i} + (1-\beta_{k+1} ) \sum_{i=1}^p\lambda_i^{1/2}{\mathbf{w}_i}'^T\circ\boldsymbol{\pi_k}^T\mathbf{q}'_{k+1},\,k=1,...,K-1,\,i=1,...,p  \\ \nonumber
&{\mathbf{u}'_{0}}&=&\beta_0 \mathbf{L}_0^T \mathbf{q}'_{0}+\mathbf{x}_0 \\\nonumber
&\boldsymbol{\alpha}'_{0} &=&\boldsymbol{\alpha}'_{0} +(1-\beta_{0} ) \sum_{i=1}^p\lambda_i^{1/2}{\mathbf{w}_i}'^T\circ\boldsymbol{\pi_0}^T\mathbf{q}'_{k+1},\,i=1,...,p   \\\nonumber
&\mathbf{s}^i&=&\mathbf{C}^{1/2}{\boldsymbol{\alpha}^i}',\,i=1,...,p \\ \nonumber
&\mathbf{L}^T \mathbf{x} &=& \left ( \mathbf{u'}_{0},{\mathbf{s}^i}',\, i=1,...,p \right ),
\end{align}

Algorithms (\ref{eq11}) and (\ref{eq12}) allow one to perform the inner-loop minimization of incremental 4D-Var, at the end of which an increment ($\delta \mathbf{x}_a$) is produced. Each outer-loop iteration consists in integrating a non-linear trajectory starting from the previous first-guess augmented by the new increment. The algorithm to update the cost function (\ref{eq8}) at each outer iteration $j+1$ is as follows:
\begin{align}
\label{eq13}
\forall k,\, & \mathbf{x}_k^{{j+1}} & = & \mathbf{x}_k^{{j}}+({\delta \mathbf{x}_k^a})^{{j}} \\ \nonumber
&                      & = & \mathbf{x}_k^{{j}}+\mathbf{L}_k({\mathbf{v}_k^a})^{{j}} \\ \nonumber
&  {\mathbf{d}_{k}^o}^{{j+1}}   & = & \mathbf{y}_{k}-h_{k}(\mathbf{x}_k^{{j+1}}) \\\nonumber
&  {\mathbf{d}_{k}^b}^{{j+1}}   & = & \mathbf{x}_k^b-\mathbf{x}_k^{{j+1}}
\end{align}
Note that in practice the implementation of this ensemble-variational method requires to consider $K$ subwindows, so that the operators $h_k$ include both the model integration and the application of the observational operator within each subwindow $k$. From (\ref{eq13}) one sees that once the increment $({\delta \mathbf{x}_k^a})^{{j}}$ has been evaluated the integration of each subwindow trajectory $h_{k}(\mathbf{x}_k^{{j+1}})$ (and thus the evolution of the first-guess departures ${\mathbf{d}_{k}^o}^{{j+1}}$) can be performed in parallel.



\subsubsection{Error propagation}
In the current IFS 4D-Var, posterior sampling of the pdf and posterior error covariance propagation to the next assimilation cycle is performed through the EDA, an EnKF-like Monte-Carlo approximation based on an ensemble of perturbed analysis \cite{bonavita2012}. 
Our ensemble-variational method, by defining a 4D control vector throughout the assimilation window, would allow one to construct an approximation of the posterior error covariance matrix at final time $t_f$ (i.e., at the initial time of the next assimilation window) as a by-product of the minimization using Hessian information \cite{bousserez2018}\cite{auligne2016}, therefore avoiding costly additional (albeit parallel) analysis computations. More specifically, at the end of the incremental 4D-Var minimization, the following formula can be used to approximate the square-root of the posterior error covariance matrix \cite{bousserez2018}:
\begin{align}
\label{eq16}
\mathbf{S}_a \approx\widehat{\mathbf{S}_a} \equiv \mathbf{L}\left( \sum_{i=1}^{q} \left ( (1+\widehat{\lambda_i})^{1/2}-1 \right) \widehat{\mathbf{v}_i}\widehat{\mathbf{v}_i}^T + \mathbf{I} \right),
\end{align}
where $\{\widehat{\lambda_i},\widehat{\mathbf{v}_i},\, i=1,...,q\}$ are, in the case of a Conjugate-Gradient (CG) optimization, Ritz pairs approximations of the $q$ largest largest eigenvalues and eigenvectors. Note that $\{\widehat{\mathbf{v}_i},\, i=1,...,q\}$ and $\mathbf{L}$ are 4D vectors and operator, respectively, defined at all time indices $k$. A posterior sampling method is obtained by applying the square-root operator $\mathbf{S}_a$ to a set of independent random 4D input vectors $\{\boldsymbol{\epsilon}_j,\, j=1,...,p\}$, i.e:
\begin{align}
\label{eq17}
\forall j=1,...,p,\, \delta \mathbf{x}^{a,j}&=\widehat{\mathbf{S}_a}\boldsymbol{\epsilon}_j \\ \nonumber
\mathbf{x}^{a,j}_K&=\mathbf{x}^{a}_K+ \delta \mathbf{x}^{a,j}_K
\end{align}
where $\mathbf{x}^{a}_K$ represents the analysis at final time of the assimilation window, and $\{\delta \mathbf{x}^{a,j}_K\, j=1,...,p\}$ are the associated posterior perturbations. Those perturbed states are then used to define the ensemble-based background error covariance matrix for the next assimilation window, as defined in (\ref{eq11}), i.e.:
\begin{align}
\label{eq18}
\forall j=1,...,p,\, k=1,...,K-1,\,  \mathbf{x}^{b,j}_0&={\mathbf{x}^{a,j}_K}^- \\ \nonumber
\mathbf{w}_k^j&=\mathbf{m}_{0 \rightarrow k}\mathbf{x}^{b,j}_0 \\ \nonumber
{\mathbf{w}_k^j}'&=1/\sqrt{p-1}(\mathbf{w}_k^j-\overline{\mathbf{w}_k}),
\end{align}
where $-$ denotes the previous assimilation cycle, $p$ is the number of samples used to generate the perturbations $\{{\mathbf{w}_k^j}'\}$, and $\mathbf{m}_{0 \rightarrow k}$ represents the non-linear model integration from $t_0$ to $t_k$. Combining (\ref{eq11}) and (\ref{eq16}), one sees that the computation of each sample $\delta {\mathbf{x}}^{a,j}$ requires one integration of the TL model, and that all $p$ samples can be computed in parallel. Thus, the cost of generating $p$ samples of the posterior pdf is significantly lower than that of an EDA, for which $p$ 4D-Var minimizations would need to be performed.

\subsubsection{4D localization of the sample error covariances}
\label{sec:loc}
Due to the low-rank nature of the ensemble-based covariance matrix $\mathbf{B}_\text{ens}$ compared to the dimension of the inverse problem, sampling noise is expected to significantly degrade the quality of the estimates. Several techniques have been designed to address this issue in ensemble-based data assimilation, from ad hoc prescriptions of maximum spatial correlation radiuses to more sophisticated statistical approaches. Here we will use recent developments in sample covariance filtering by \cite{menetrier2015linear}, which are implemented in the BUMP ("Background error on Unstructured Mesh Package") software. This tool can both diagnose 4D correlation lengths of $\mathbf{C}$ globally and construct the associated NICAS ("Normalized Interpolated Convolution from an Adaptive Subgrid") smoother. The BUMP software is being implemented in OOPS ("Object Oriented Prediction System"), which is the C$^{++}$ code that will be used as the assimilation layer of the future operational IFS.

\subsection{A Kalman smoother for CO$_2$ source optimization}
\label{sec:kalman_smoother}
The current version of the IFS 4D-Var system consists of a Kalman filter-like algorithm, wherein the control variables are optimized only in the current 12-hour window and the error statistics are propagated forward in time. As mentioned in Sec. \ref{sec:IFS_const}, the CO$_2$ inversion prototype requires flexibility with regard to the assimilation window length, so that CO$_2$ emissions spanning weekly to monthly time-scale can be jointly optimized using observations covering the entire period of interest. 
In this section a methodology to implement a Kalman smoother algorithm is proposed. It relies on extending to past and future windows the current 12-hour 4D-Var assimilation window for CO$_2$ source estimation. In an operational context such as the one envisioned for the prototype, such an adaptation needs to be non-intrusive (i.e., minimize changes in the current 4D-Var algorithm implementation) and represent a small additional cost to the optimization procedure. The latter constraint prevents the use of a full online adjoint-based optimization for long 4D-Var windows, i.e., only the CO2 emissions and the associated tracer transport processes can be considered outisde the short 12-hour online window. Note that non-linearities from meteorological processes could also jeopardize the optimization efficiency due the presence of multiple local minima. 


\subsubsection{Ensemble-based transport Jacobians}
\label{sec:jacobian}
One way to overcome the computational bottleneck associated with long assimilation windows is to consider an ensemble-based approximation of the Jacobian of the tracer transport. At the end of each 4D-Var assimilation cycle (window $k$), an ensemble of forward CO$_2$ transport simulations associated with posterior emission perturbations starting from the same analysis fields is generated. The 3D CO$_2$ concentrations obtained at the end of the online assimilation window $k$ are archived. During the next assimilation cycle (window $k+1$), the control vector increment consists of only the CO$_2$ emissions in windows $j<k+1$ and all variables (i.e., meteorological fields, chemical tracers) in window $k+1$. The TL propagation of the emission increments in windows $j<k+1$ is performed using a least-square approximation of the Jacobian based on the archived ensemble of CO2 concentrations and associated emissions perturbations (i.e., similar to a finite-difference evaluation of the Jacobian, but in a basis defined by the posterior emission perturbations). At the end of window $k$, the propagation of the CO$_2$ concentration increment to the end of window $k+1$ is performed using the standard TL solver. The backward adjoint integration follows the same chain-rule principle but using the transpose of the least-square Jacobian approximation. The linearity of the transport processes allows one to reuse the ensemble-based Jacobian approximation at each outer-loop of the minimization procedure and for all subsequent assimilation windows.  Figure \ref{fig:fig1} shows a schematic for the near-real time inversion system. Note that the quality of the sampling to construct the Jacobian is critical to the success of this ensemble-based approach. Investigations are underway to build a Jacobian sampling approach that would maximize the information content of the inversion accross multiple windows. Two natural choices, although not necessarily optimal, are to sample: 1) the prior emissions using the square-root of the prior error covariance matrix (i.e, $\mathbf{L}$); 2) the posterior emissions using the square-root of the posterior error covariance matrix provided by (\ref{eq16}).

\subsubsection{Ensemble-based transport error covariance}
\label{sec:transport_err}
\begin{figure}
\centering
\includegraphics[width=30pc]{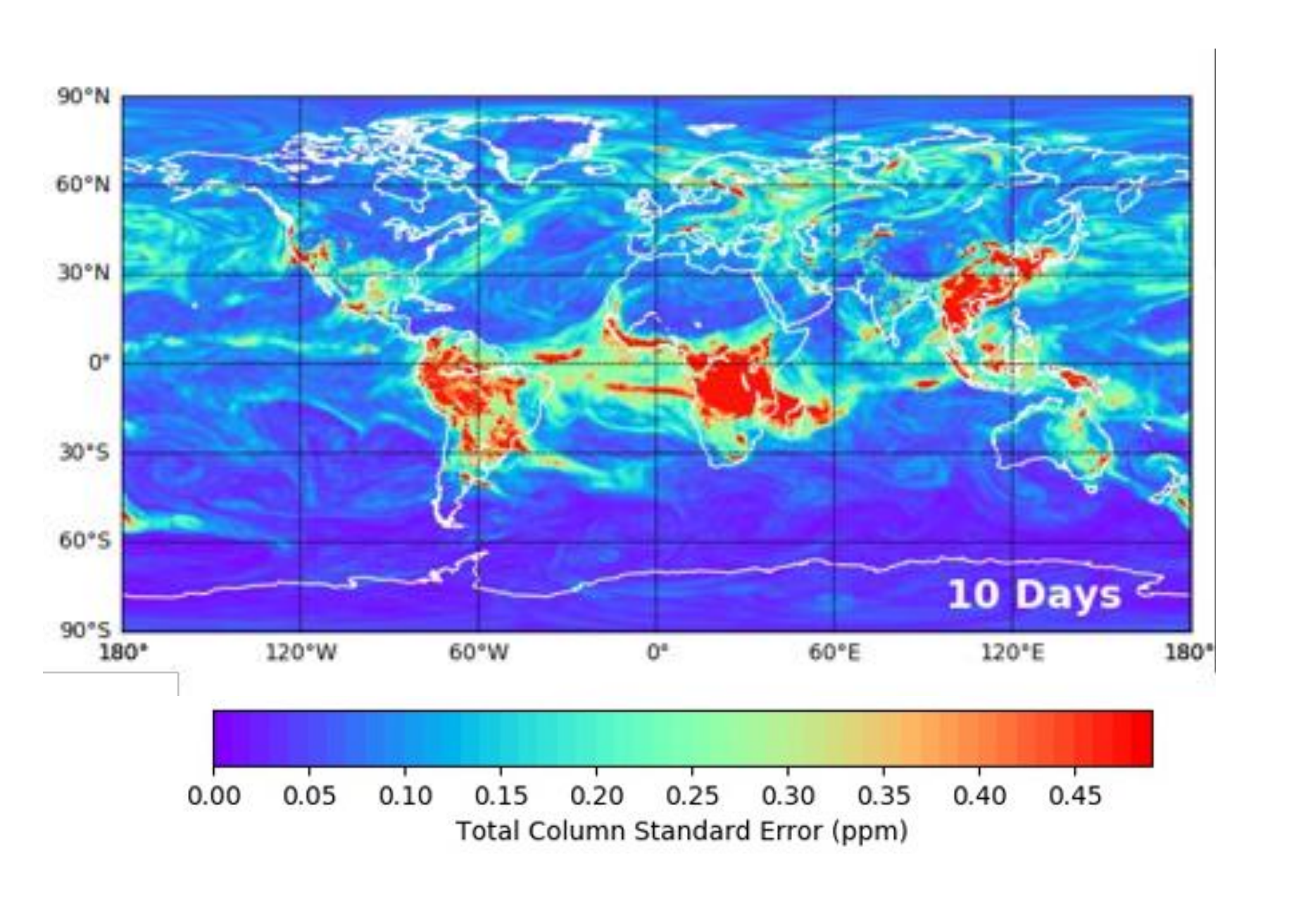}
  \caption{Standard error deviation of the IFS CO$_2$ columns (XCO$_2$) based on an ensemble of 50 perturbed CO$_2$ forward model simulations driven by the EDA after 10 days of integration.}
  \label{fig:fig3}
\end{figure}

For the re-analysis product (see Sec. \ref{sec:reanalysis}), observations spanning a long window (e.g., weeks to months) are all assimilated at once. Therefore, in the subwindows where ensemble-based offline approximations are used (i.e., outside the online 12-hour window), the transport error associated with the modeled CO$_2$ concentrations has to be explicitly accounted for in the observational error covariance matrix $\mathbf{R}$. To this aim, an ensemble of IFS forward CO$_2$ simulations starting with the same CO$_2$ prior emissions but using different meteorological fields from an EDA will be used to estimate the transport error covariances. The raw sample estimate obtained through this approach will then be localized in space and time using the BUMP software as described in Sec. \ref{sec:loc}. This software can efficiently handle error covariance filtering and modeling on irregular grids, which is key in the case of error statistics defined in observation space. Figure \ref{fig:fig3} represents the standard error deviation of the IFS integrated CO$_2$ columns (XCO$_2$) based on an ensemble of 50 perturbed forward model simulations after 10 days of integration. The impact of transport error on the CO$_2$ columns is most pronounced over regions with highest anthropogenic and biogenic emissions (i.e., eastern Asia, central Africa and South America), where emission gradients are also the strongest. 

\subsection{Near-real time product}
\label{sec:near_realtime}
\begin{figure}
\centering
\includegraphics[width=30pc]{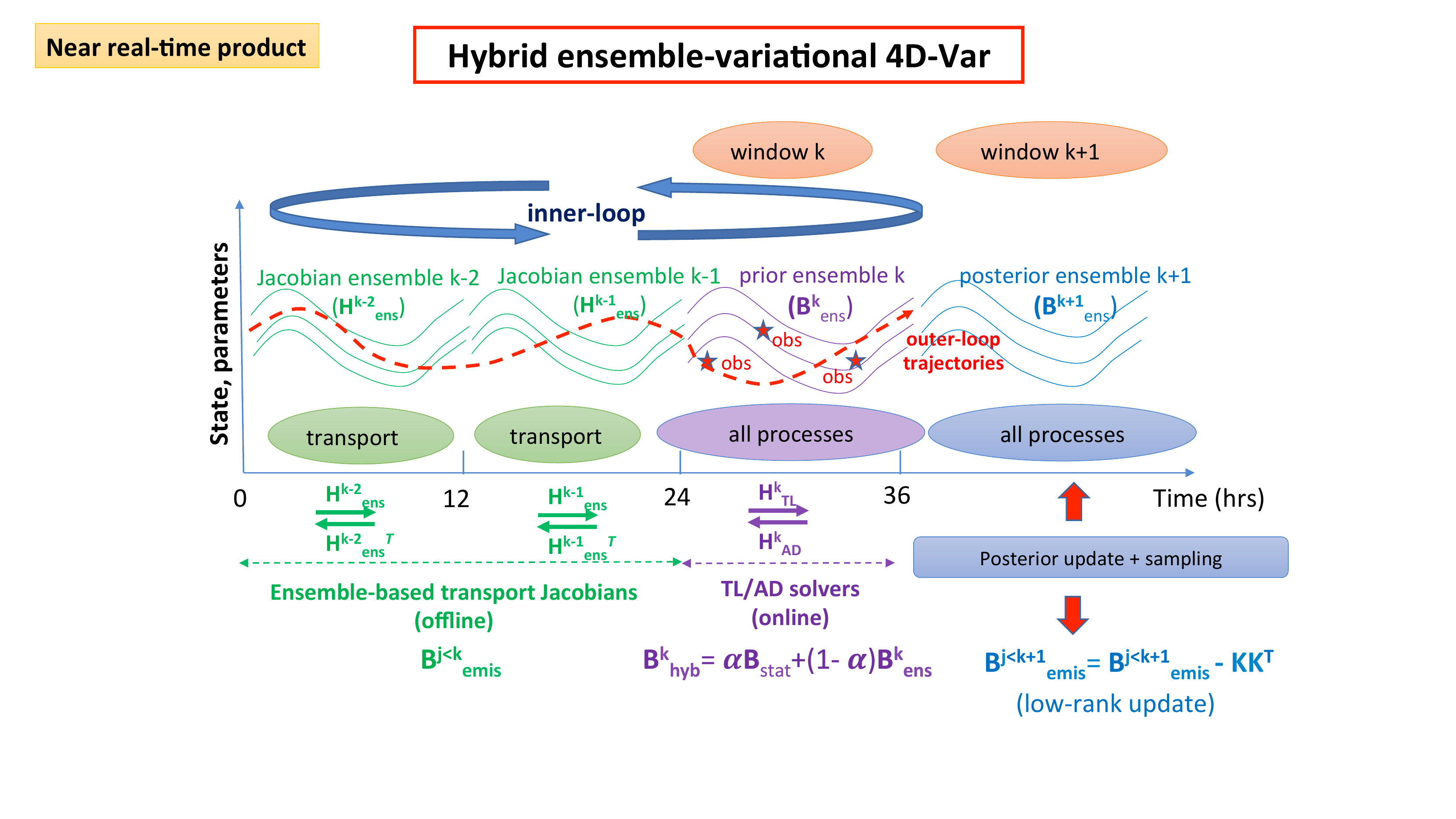}
  \caption{Schematic of the near-real time product for the prototype.}
  \label{fig:fig1}
\end{figure}

The near-real time configuration proposed is adapted to the need of an operational fast delivery product. Figure \ref{fig:fig1} represents a schematic of the system. For the short online window, the control vector includes both the CO$_2$ emission scaling factors and the assimilated meteorological fields (possibly the reactive species emission scaling factors and concentrations as well), while all the processes (i.e., tracer transport and prognostic atmospheric state equations) are included in the forward model and TL/AD integrations. Note that in the ensemble-variational system described in Sec.\ref{sec:hyb_B} the 4D atmospheric state is optimized, accounting for model errors. Here errors in CO$_2$ modeling not associated with transport uncertainty (e.g., from chemical reactions) are neglected, and therefore only the CO$_2$ emissions are optimized. This is in contrast with the optimization of reactive trace gas (e.g., NO$_x$, CO), for which the associated 4D concentrations would be jointly optimized with the emissions (i.e., model error in the chemistry is accounted for). Note that optimizing the CO$_2$ emissions only also enforces CO$_2$ mass conservation. The short online window is linked to past assimilation windows through the ensemble-based Jacobians $\mathbf{H}_\text{ens}^j$. By applying the chain rule and combining those Jacobians with the TL and AD solver integrations, scaling factor increments can be propagated forward and backward in time at each iteration across the 4D-Var subwindows. A posterior ensemble at initial time is then generated using the posterior error square-root provided by (\ref{eq16}), and each ensemble member propagated by the forward model to the initial time of the next assimilation window. 

\subsection{Re-analysis product}
\label{sec:reanalysis}
\begin{figure}
\centering
\includegraphics[width=30pc]{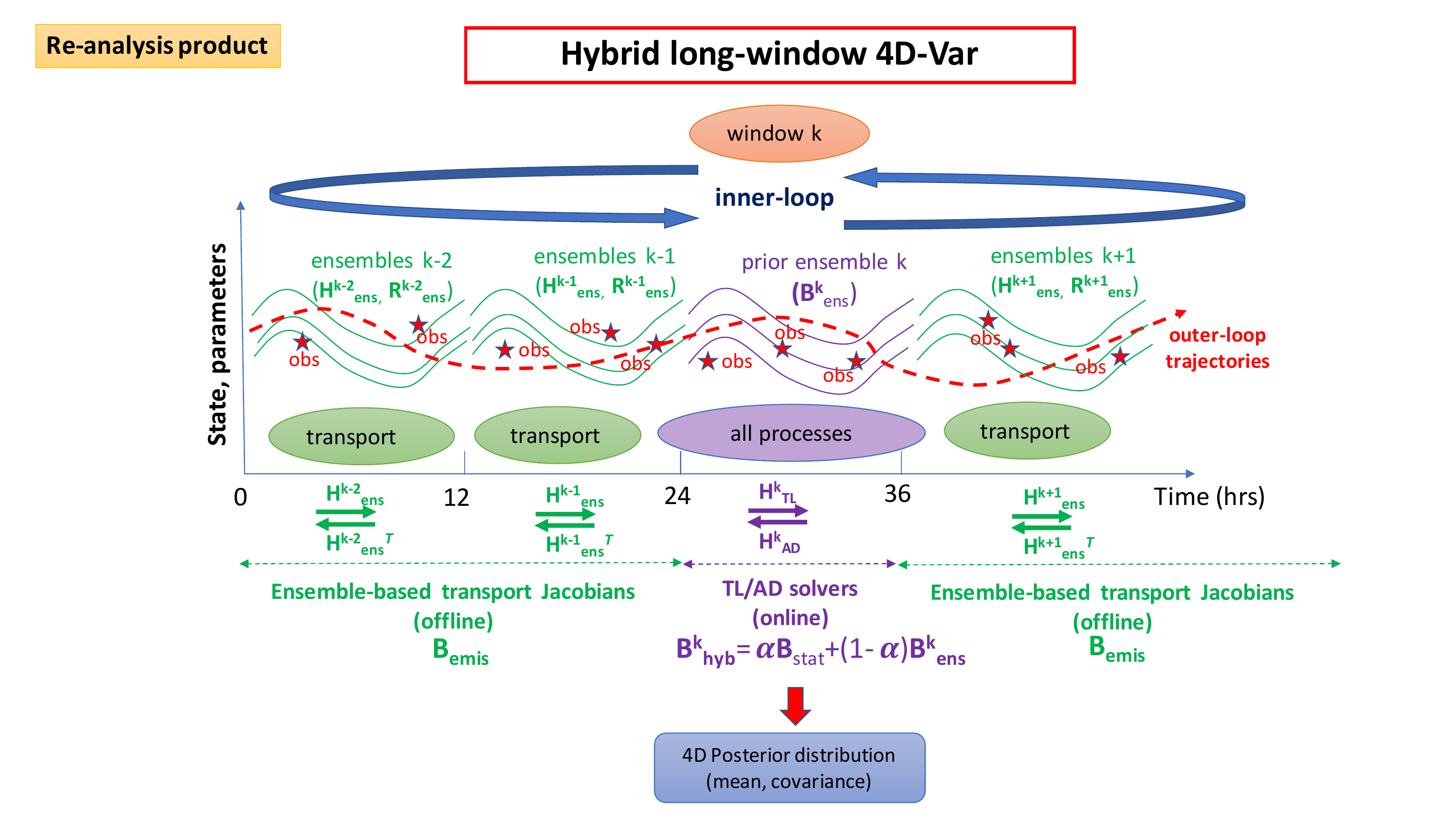}
  \caption{Schematic of the re-analysis product for the prototype.}
  \label{fig:fig2}
\end{figure}

The purpose of the re-analysis configuration is to provide a better estimate of the CO$_2$ fluxes that accounts for all observations in current, future and past assimilation windows. In addition, this configuration facilitates the integration of multi-model posterior products in the global IFS-driven prototype (see Sec. \ref{sec:multi}). Since the online short-windows are beneficial to account for multiple co-emitted reactive species and correct for model transport error of CO$_2$ plumes in the vinicity of emission hotspots, the propose methodology consists in performing, for each short-window, a long-window 4D-Var re-analysis "centered" around that window. Figure \ref{fig:fig2} represents a schematic of the method. The principle is similar to the one proposed for the near-real time product (see Sec. \ref{sec:near_realtime}), except that only one minimization is performed (i.e., there is no cycling of the DA) to assimilate all observations spanning the period of interest at once. As a result, for the offline (ensemble-based) windows, the transport error statistics associated with the observations have to be explicitly represented, since they are no longer implicitly accounted for as in the online assimilation configuration. The transport error statistics will be estimated and represented as described in Sec. \ref{sec:transport_err}.

\subsection{Tuning of the $\mathbf{B}$ matrix for chemical emissions}
Spatial and temporal error covariances in the prior emissions are highly uncertain, yet accurately representing them is crucial to the success of the inversion method. In contrast to transport error estimations, which can be partially inferred from an ensemble-based approach wherein model parameters are perturbed (see Sec.\ref{sec:transport_err}), errors in the bottom-up emission inventories are more difficult to sample. Typically, only error standard deviations for yearly budget are provided, which have then be re-scaled to define uncertainty at monthly or weekly timescale, and spatiotemporal error correlation information is missing. One way to refine the prescribed prior error covariance matrix is through optimization using atmospheric CO$_2$ observations. Due to the high-dimensional nature of the error covariance for global grid-scale inversions, this matrix is never explicitly formed and has to be modeled using operators. In the two following sections we described the model used for the $\mathbf{B}$ matrix as well as a method to optimize its parameters.

\subsubsection{Wavelet-J$_b$}
An efficient model for the error correlations in the $\mathbf{B}$ matrix can be obtained through a spectral representation \cite{courtier1998}. This approach provides a compact representation of the matrix using a number of spectral functions that is much smaller than the initial dimension of the grid-scale emission space. However, this method does not allow for spatial variation of the error correlation length characteristics. Spatial error correlation structures in the prior emission inventories are expected to be highly heterogeneous geographically due to the complex spatial distribution of the different emission sectors and vegetation types. Therefore, we propose to use a so-called wavelet-J$_b$ approach, wherein spatial variations of the error correlation structures can be modeled \cite{fisher2006wavelet}. In practice, this consists in performing a convolution between a basis of wavelet functions and the error correlation field of interest, which is then reconstructed using the wavelet decomposition. The reader can refer to \cite{fisher2006wavelet} for more details on the method. 

\subsubsection{Maximum-likelihood optimization}
In this section we describe an approach to tune the hyperparameters of the prior error covariance matrix for chemical emissions. The method is based on the maximum-likelihood principle \cite{wu2013}, wherein the maximum of the following conditional pdf is sought:
\begin{align}
\label{eq21}
p(\boldsymbol{\theta}|\boldsymbol{y})=\frac{p(\boldsymbol{y}|\boldsymbol{\theta})p(\boldsymbol{\theta})}{p(\boldsymbol{y})},
\end{align}
where $\boldsymbol{y}$ represents the observation vector and $\boldsymbol{\theta}$ the vector of hyperparameters for the $\mathbf{B}$ matrix.

Assuming a Gaussian pdf for the likelihood and prior, maximizing (\ref{eq21}) leads to minimizing the following cost function:
\begin{align}
\label{eq22}
\mathcal{L}(\boldsymbol{\theta})=\frac{1}{2}\ln |\mathbf{D}_\theta|+\frac{1}{2}(\mathbf{y}-\mathbf{H}\mathbf{x}_b)^T\mathbf{D}_\theta^{-1}(\mathbf{y}-\mathbf{H}\mathbf{x}_b),
\end{align}
where $\mathbf{H}$ is the linear transport model operator and $\mathbf{D}_\theta=\mathbf{R}+\mathbf{H}\mathbf{B}_\theta\mathbf{H}^T$ is the so-called covariance matrix of innovation statistics.
In the Desroziers approach \cite{desroziers2001} \cite{chapnik2006}, only a multiplicative scaling factor is optimized, i.e., the correlation structures of $\mathbf{B}$ are assumed to be known, such that the true error covariance matrix can be written $\mathbf{B}^t=\alpha\, \mathbf{B}$, where $\alpha$ is the parameter to be optimized. The optimization then relies on a fixed-point algorithm wherein a complete 4D-Var minimization has to be carried out at each iteration. In the case of prior emission error covariances, the correlation structures are initially very uncertain, so that those quantities need to be resolved by the optimization scheme. In the case of a wavelet-based $\mathbf{B}$ matrix, this can be achieved by optimizing the coefficients associated with each wavelet basis function. In practice, a gradient-based optimization algorithm (e.g., the BFGS algorithm) can be used to find the minimum of (\ref{eq22}). A computationally tractable method relies on the ability to efficiently evaluate the adjoint of the transport model ($\mathbf{H}$) as well as the matrix of innovation statistics $\mathbf{D}_\theta$. In our approach, the Jacobian $\mathbf{H}$ will be approximated using the ensemble technique described in Sec. \ref{sec:jacobian}. Since the tuning of the parameters $\boldsymbol{\theta}$ will be performed offline, a large sample may be used to span a sufficiently large subspace. Moreover, the approximation of the transport error covariance matrix $\mathbf{R}$ will be constructed using the ensemble-based method described in Sec. \ref{sec:transport_err}. A 4D localization tehnique (see Sec. \ref{sec:loc}) will ensure proper filtering of the transport error sampling noise. Formally, the following approximations will be used in the minimization of  (\ref{eq22}):

\begin{align}
\label{eq23}
&\mathbf{D}_\theta&= &\mathbf{R}_\text{ens}+\mathbf{H}_\text{ens}\mathbf{B}_\theta\mathbf{H}_\text{ens}^T \\ \nonumber
&\mathbf{R}_\text{ens}&= &\mathbf{C}\circ\mathbf{X}_{\mathbf{R}}\mathbf{X}_{\mathbf{R}}^T\\ \nonumber
&\mathbf{H}_\text{ens} &=& (\mathbf{H}\mathbf{X}_\text{H})\mathbf{X}_\text{H}^+ \\ \nonumber
&\mathbf{X}_\text{H}& = &\mathbf{U} \\ \nonumber
&\mathbf{L}_{\theta_0}^T\mathbf{H}^T\mathbf{R}^{-1}\mathbf{H}\mathbf{L}_{\theta_0}&=&\mathbf{U} \mathbf{\Lambda}\mathbf{U}^T, \nonumber
\end{align}
where $\mathbf{B}_{\theta_0}$ represent the error covariance matrix associated with the first guess value of the parameters $\theta$ (i.e., the starting point of the minimization), an $\mathbf{U} \mathbf{\Lambda}\mathbf{U}^T$ represents the eigendecomposition of the prior-preconditioned Hessian associated with $\boldsymbol{\theta}_0$. The sampling ($\mathbf{U}$) is designed so as to maximize (in a statistical sense) the information content of the resulting approximated transport Jacobian.

The tuning of the spatial error covariances for the emission of individual chemical species can be performed jointly with the optimization of inter-species error correlations in wavelet space, in which case those correlations are part of the control vector $\boldsymbol{\theta}$. Similarly, the optimization of temporal error correlations will be considered by including those in the wavelet control vector. In practice, atmospheric measurements of CO$_2$, NO$_2$ and CO will be jointly used to estimate the cross-species error correlations, which in turn will provide further constraints on CO$_2$ source attribution (e.g., biospheric fluxes, sectorial attribution).

\subsection{Multi-model inversion}
\label{sec:multi}
Integrating heterogeneous inversion products (i.e., global to regional to local posterior emissions ) into one single global inversion system would provide advantages from both a practical (readability) and theoretical (statistical consistency) standpoint. Here we describe a method that would allow one to merge global, regional and local posterior CO$_2$ fluxes generated by different project partners into the global IFS inversion system. This approach consists in considering each external inversion product as an observation to be assimilated in the re-analysis IFS hybrid ensemble-variational system described in Sec. \ref{sec:reanalysis}. The observational operator that maps the true emissions $\mathbf{x}$ to the inverted emissions $\mathbf{x}'$ is the averaging kernel matrix $\mathbf{A}$ of the inversion algorithm. In particular, one has:
\begin{align}
\label{eq24}
&\mathbf{x}'&=&\mathbf{x}_b+\mathbf{A}(\mathbf{x}-\mathbf{x}_b)+\mathbf{K}\boldsymbol{\epsilon} \\ \nonumber
&\mathbf{K}&=&\mathbf{B}\mathbf{H}^T(\mathbf{H}\mathbf{B}\mathbf{H}^T+\mathbf{R})^{-1}\\ \nonumber
&\mathbf{A}&=&\mathbf{KH}\\\nonumber
&\boldsymbol{\epsilon}&=&\mathcal{N}(\boldsymbol{0},\mathbf{R}_\epsilon),\nonumber
\end{align}
where $\mathbf{x}_b$, $\mathbf{K}$ and $\mathbf{H}$ correspond to the quantities used in the external inversion product, while $\boldsymbol{\epsilon}$ and its covariance $\mathbf{R}_\epsilon$ correspond to the true observational errors that include both measurement and forward model (e.g., transport) uncertainties.

The likelihood function associated with one assimilated posterior emission product is defined as follows: 
\begin{align}
\label{eq25}
&p(\mathbf{x}'|\mathbf{x})& = &\mathcal{N}(\mathbf{x}_b+\mathbf{A}(\mathbf{x}-\mathbf{x}_b),\mathbf{S}) \\ \nonumber
&\mathbf{S}&=&\mathbf{K}\mathbf{R}_\epsilon\mathbf{K}^T
\end{align}

In practice, using the approximation of the Kalman gain matrix $\mathbf{K}$ provided in (\ref{eq27}), one can compute $\mathbf{S}$ based on samples of the observational error $\boldsymbol{\epsilon}$, e.g., using the ensemble method described in Sec. \ref{sec:transport_err}.
In a variational context, the gain matrix $\mathbf{K}$ and $\mathbf{H}$ are generally not available, since the solution is computed iteratively using TL and AD model integrations. Approximations of $\mathbf{A}$ and $\mathbf{K}$ can nevertheless be obtained based on sample estimates of the prior and posterior pdfs. More specifically, the following expression can be used to estimate the averaging kernel matrix:
\begin{align}
\label{eq26}
&\mathbf{A}&=&\mathbf{I}-\mathbf{P}^a\mathbf{B}^{-1} \\ \nonumber
&&\approx&\mathbf{I}-{\mathbf{X}^a}'{{\mathbf{X}^a}'}^T \left ( {\mathbf{X}^b}'{{\mathbf{X}^b}'}^T \right )^{-1} ,
\end{align}
where $\mathbf{X}'=\frac{1}{\sqrt{N-1}}\left( \mathbf{X}-\overline{\mathbf{X}}\right )$ represents the matrix column of sample deviations from the mean associated with either the prior (superscript $b$) or posterior (superscript $a$) of the external inversion product. Similarly, one can derive an approximation of the Kalman gain matrix $\mathbf{K}$ as follows:
\begin{align}
\label{eq27}
&\mathbf{K} \approx {\mathbf{X}^b}'(\mathbf{H}{\mathbf{X}^b}')^T\left (\mathbf{H}{\mathbf{X}^b}'(\mathbf{H}{\mathbf{X}^b}')^T+\mathbf{R} \right )^{-1},
\end{align}
where $\mathbf{R}$ can be provided either as a matrix or via a sample estimate of the observational error (e.g., if correlations are present).

Finally, when assimilating an external posterior emission into the global IFS system, if observations used in the external inversion are used in the global IFS inversion as well, error covariances between those observations and the assimilated posterior emissions need to be evaluated. An approximation of those error covariances can be obtained as follows:
\begin{align}
\label{eq28}
\mathrm{cov}(\mathbf{x}',\mathbf{y})\approx{\mathbf{X}^a}'^T{\mathbf{Y}}',
\end{align}
where $\mathbf{y}$ is an observation vector assimilated in both the external posterior emission product and the global IFS re-analysis system, and ${\mathbf{Y}}'$ is the matrix column of sample deviations from the mean associated with the observation $\mathbf{y}$. In practice, along with the posterior mean emissions to be assimilated, the sample estimates ${\mathbf{X}^b}'$, ${\mathbf{X}^a}'$ and associated ${\mathbf{Y}}'$ would need to be provided.
\section{Conclusion}
A hybrid ensemble-variational assimilation system that can implement chemical source inversion capabilities in the IFS has been presented. The main characteristics of the proposed methodology and most noticeable differences from the current 4D-Var algorithms can be summarized as follows:
\begin{itemize}
    \item Model random errors can be included in the variational optimization using, e.g., an ensemble of forward model simulations with Stochastically Perturbed Physics Tendencies (SPPT) to construct a 4D generalization of the hybrid $\mathbf{B}$ formulation. This approach is theoretically more robust than the current strong-constrained 4D-Var framework, which does not account for model errors in the analysis.
     \item Unlike with the ensemble-variational method, a 4D localization of the ensemble-based ($\mathbf{B}_\text{ens}$) matrix is partially achieved through the combination of the ensemble increment with a full-rank increment propagated with the TL/AD models.
    \item Processes (e.g., chemical reactions) that are not accounted for and/or are modeled at coarser resolution in the TL/AD integrations can be included in the propagated ensemble, providing a computationally efficient way of including more complexity in the increment propagation while maintaining low integration cost for the TL/AD models.
    \item Joint inversion of CO$_2$ and co-emitted species (e.g., CO, NO$_2$) can provide additional constraints on the source attribution and help disentangle the anthropogenic and biogenic components of the signal.
    \item Ensemble-based approximations of transport Jacobians can be used to extend the current 12-hour 4D-Var assimilation window and enable long-window 4D-Var for GHG source inversions.
    \item External posterior flux products, obtained using different models and domains (from global to regional to local), can be assimilated as observations in the global IFS-driven source inversion system provided samples of the posterior and prior distributions are available. Such quantities are by-product of EnKF-like algorithms, while in variational approaches Monte-Carlo techniques can be used to generate them. 
   \end{itemize}
   The proposed methodology will be implemented in the new OOPS data assimilation layer. A follow-up document will provide the detailed implementation steps for the prototype, including computational requirements, workflow and associated timeline.
\newpage
\bibliographystyle{plain}
\bibliography{mybib.bib}{}
\end{document}